\documentclass[a4paper,12pt]{article}

\usepackage[english]{babel}
\usepackage{dsfont}
\usepackage[utf8]{inputenc}
\usepackage{amsthm}
\usepackage[dvipsnames]{xcolor}
\usepackage{tikz}
\usetikzlibrary{automata,positioning}
\usepackage{amssymb}
\usepackage{mathtools}
\usepackage{xparse,graphicx}
\usepackage[skip=0.5ex]{subcaption}
\usepackage{bm}
\usepackage{bbm}
\usepackage{multirow,array}
\usepackage{float}
\usepackage{titlesec}
\usepackage{titling}
\usepackage{lipsum}
\usepackage[colorinlistoftodos]{todonotes}
\usepackage[longnamesfirst]{natbib}
\usepackage[margin=1.25in]{geometry}
\usepackage{setspace}
\usepackage[strict]{chngpage}
\usepackage[titletoc,title]{appendix}
\usepackage{pgfplots}
\usepgfplotslibrary{groupplots}
\pgfplotsset{compat=newest}
\usepackage{caption}
\usepackage{subcaption}
\usepackage{dirtytalk}
\usepackage{nicefrac}
\usepackage{booktabs}

\usepackage{hyperref}
\definecolor{linkcolour}{rgb}{0,0.2,0.6} 
\hypersetup{
    colorlinks=true,
    linkcolor=black,
    citecolor=linkcolour,
    filecolor=black,
    urlcolor= linkcolour
}

\usepackage{enumitem}
\usepackage{bigints}
\usepackage{comment}

\usepackage{natbib}
\usepackage{hyperref}

\newtheorem{lemma}{Lemma}

\newtheorem{remark}{Remark}

\newtheorem{proposition}{Proposition}

\allowdisplaybreaks

\makeatletter
\newcommand{\vast}{\bBigg@{3.5}}
\newcommand{\Vast}{\bBigg@{5}}
\makeatother

\definecolor{RoyalBlue}{RGB}{0, 76, 153}
\definecolor{C_purple}{RGB}{153, 0, 153}
\definecolor{N_Blue}{RGB}{51, 153, 255}
\definecolor{A_green}{RGB}{0, 153, 0}
\definecolor{An_orange}{RGB}{255,128,0}

\definecolor{myPurple}{rgb}{0.6,0.1,0.6}
\definecolor{myNavy}{rgb}{0.0,0.2,0.4}
\definecolor{myLightBlue}{rgb}{0.2,0.6,0.9}
\definecolor{myGreen}{rgb}{0.0,0.4,0.0}
\definecolor{myOrange}{rgb}{0.9,0.4,0.1}

\setlist{itemsep=0.2pt,topsep=1.6pt}

\title{\LARGE An Evolutionary Analysis of Narrative Selection\thanks{We thank Flavio Contrada, Pietro Dindo, Daniele Giachini, Paolo Pin, Alessandro Stringhi, Anastas Tenev, Alpen Yasar, the participants at the ASSET 2024 meeting, the University of Siena internal seminar, and the first meeting of the Economic Tuscan Theorists, as well as two anonymous referees, for their valuable comments and suggestions. \newline [\href{https://github.com/rrozzi-econ/An-evolutionary-analysis-on-narrative-selection/tree/main}{Repository folder}]}}

\author{Federico Innocenti\thanks{Department of Economics, University of Konstanz; \href{f.innocenti93@gmail.com}{\nolinkurl{f.innocenti93@gmail.com}}.} \and Roberto Rozzi\thanks{Department of Psychology,
Royal Holloway University of London, email:\\ 
\href{Roberto.rozzi@rhul.ac.uk}{\nolinkurl{Roberto.rozzi@rhul.ac.uk}}.}}

\date{\today}

\begin{document}

\maketitle

\begin{abstract}
    We compare the evolutionary selection of agents who employ different heuristics to select the narratives through which they interpret new information. Agents' survival depends on the accuracy of their beliefs. Agents who choose narratives to align with the population's average belief have higher survival rates than those who use alternative heuristics. Conforming reduces an agent's belief variance but inherits the bias of the average belief, so it pays when the latter is close to the state's probability. Evolutionary selection helps maintain this proximity by eliminating inaccurate agents. The other heuristics survive as well, with their shares depending on the degree of uncertainty: agents who tend to form mild beliefs survive more when uncertainty is high, whereas those who tend to form extreme beliefs survive more when uncertainty is low. Conformism is usually considered detrimental, as it discards private information. In our model, conformist agents read their private information in light of the public information, and the consensus they anchor to consists of the agents that survived evolutionary selection, so the cost of conformism is limited.

\noindent \\
\vspace{-0.75cm}
\vspace{0in}
\end{abstract}

\newpage

\section{Introduction}\label{sec_intro}

How do people process new information? This question lies at the heart of numerous socio-economic studies \citep{enke2024cognitive}. A substantial body of research indicates that people exhibit confirmation bias when updating their beliefs: they attribute greater weight to information that aligns with their preexisting views. Recent works in this area assume that people actively select narratives that best fit their current beliefs \citep{schwartzstein2021using}. For example, in financial markets, investors receive news about the same firm (an earnings report, a price movement, a piece of media coverage), but disagree about what it means: noise to one, decisive to another. This disagreement concerns not the information itself but how much to trust it. Investors who misread the evidence predict the firm's performance worse than their peers and are more likely to exit the market. This raises a natural question: which way of interpreting new information is most likely to survive market selection?

In this paper, we study an evolutionary model that encompasses several types of agents using different heuristics to choose a narrative (that is, a particular way to interpret new information). We show that the type most likely to survive incorporates other agents’ beliefs into its narrative choice, a heuristic we refer to as conformism. Crucially, conformist agents do not copy others: they receive their own signal, update their own belief, and let the average belief in the population guide only their interpretation of the evidence. In financial markets, this average is observable, since market indexes and prices aggregate investors' current expectations. Conformist agents' advantage stems from treating it as an anchor, which protects them from making significant errors in predicting the state of the world, especially in uncertain environments. Furthermore, we find that different types survive natural selection, and their survival chances depend on the underlying uncertainty regarding the state: types that tend to develop extreme beliefs have an evolutionary advantage when uncertainty is low, whereas those that develop mild beliefs have an advantage when uncertainty is high.

We consider a dynamic model where the state of the world is a binary random variable, drawn independently in each period, and each agent receives an independent and identically distributed binary signal about it. Agents update their beliefs using Bayes' rule, but have different information about the signal precision. Standard agents know the correct signal precision. There are four other types of agents---anti-conformist, conformist, over-confident, and under-confident---that choose between two other precision values---that is, two narratives---according to different heuristics. Over-confident agents select the narrative that best aligns with the information received among all competing interpretations \citep{schwartzstein2021using}. The fit of a narrative to the information received depends on agents' prior beliefs, which in turn depend on past information. Conformist agents care about sharing the same beliefs as their peers. We assume that conformist agents choose the narrative to match---myopically---the average prior belief in the population.\footnote{This allows us to rule out strategic considerations about peers' narrative choices.} Anti-conformist agents behave in an opposite way to conformist agents; under-confident agents behave oppositely to over-confident agents. Since agents receive independent signals and use different heuristics, they may choose different narratives.

An agent’s probability of surviving selection is strictly increasing in their accuracy in predicting the state’s actual probability (that is, in their evolutionary performance). However, agents are subject to an extreme form of system neglect \citep{massey2005detecting, kremer2011demand} because they form beliefs about the state's realization rather than about the probability governing it. In the financial markets example, investors assess a firm's current performance rather than its underlying probability of success. Those who are more accurate in predicting its prospects lose less and are thus less likely to be forced out of the market. At regular intervals, which we refer to as turnover times, some agents exit the market, and an equal number are born with random types, capturing investor entry and exit. Selection therefore operates on the heuristics themselves, as in the indirect evolutionary approach \citep{guth1992explaining,guth1995evolutionary}: the heuristics that predict the state better proliferate. We assume a random initial distribution of types in the population, and all agents are born with the same prior belief.

We characterize evolutionary performance analytically. An agent's expected error decomposes
into a bias term, which penalizes beliefs far from the state's probability, and a variance
term, which penalizes dispersed beliefs. Conforming pulls a belief toward the population’s average belief---hereafter, the anchor---reducing
its variance but inheriting whatever bias the anchor carries, so whether the trade is worth
making depends on how far the anchor lies from the state's probability. When the state is
certain, the accurate belief is extreme, and conformist agents, tied to a moderate anchor, are
at a disadvantage. When the state is uncertain, the comparison reverses. The other heuristics
deliver either an extreme or a moderate belief, neither of which follows the state's
probability. The anchor does follow it, not because the crowd is wise by assumption, but because the same selection that eliminates inaccurate agents also disciplines the anchor that conformist agents rely on. Conformist agents track the anchor by switching between heuristics: they
dampen the evidence when their belief runs ahead of the anchor and amplify it when their belief
lags behind.

We then use numerical simulations to study the long-run distribution of types in the
population for different degrees of uncertainty about the state. We find that all types
survive in the long run, although they exhibit different evolutionary performances, and thus
hold heterogeneous shares of the population. Conformist agents outperform all other types,
constituting the relative majority of the population across all specifications. The survival chances of the other types vary with the degree of state uncertainty, as our theory anticipates. On the one hand, under-confident agents exhibit better evolutionary performance when uncertainty
is high (i.e., states have similar probabilities), because the other types tend to develop
extreme beliefs, whereas under-confident agents develop milder ones. On the other hand, the
survival chances of standard and over-confident agents increase when uncertainty is low (i.e.,
one state is significantly more likely than the other), since in these situations it pays to
hold extreme beliefs. In financial markets, cautious readings of the evidence therefore survive when a firm's prospects are uncertain, and confident ones when they are not.

These results do not depend on the state being drawn independently in each period. We obtain
the same patterns when the realization of the state depends on its own past, and when the
probability of the state partly reflects the average belief in the population, e.g., because
investors' expectations feed back into the firm's performance. Conformist agents also remain the type with the largest share of the population as information precision increases and agents' misperception of information precision decreases. Lower misperception leaves them less able to reach the anchor, and their advantage narrows. Higher precision makes the anchor noisier, and thus a worse guide, but also drives the other types toward more extreme beliefs, which pays only when uncertainty is low; the conformist advantage therefore narrows with precision only in that case. The advantage of conformist agents also persists when newborn agents adopt the average belief rather than a fixed prior, thereby removing the force that keeps the anchor away from certainty and leaving selection as the only discipline on it.

Our results provide three takeaways. First, we show the importance of considering conformism in future research on narrative selection. Second, we establish the long-term coexistence of agents using different heuristics to select narratives, which is consistent with experimental evidence from cognitive psychology: individuals tend to update their beliefs based on the aggregated beliefs of others \citep{asch1951effects, bond1996culture}, and those who follow the majority coexist with those following more rigid updating rules \citep{friend1990puzzling, oktar2025aggregated}. Third, in financial markets, following the crowd is typically treated as informationally destructive: under herding, an agent's action may no longer depend on their private signal and therefore transmits none of it \citep{bikhchandani1992theory, banerjee1992simple}. Competition, in turn, is expected to eliminate those who interpret evidence poorly. Yet, conformity is pervasive among investors. Our model suggests a reason: conformism may not be a way of ignoring one's own information but of interpreting it, and the consensus to which an investor anchors is itself the product of competition among investors, thus limiting the cost of conformism.

We organize the rest of the paper as follows. In Section~\ref{sec_lit}, we review the related literature. In Section~\ref{sec_mod}, we present the theoretical model. In Section~\ref{sec_bench_naive/conf}, we characterize evolutionary performance analytically
and derive the predictions that we test in Section~\ref{sec_sim} using numerical simulations. Finally, we offer some concluding remarks in Section~\ref{sec_conc}.

\section{Literature Review}\label{sec_lit}

We contribute to the literature on narrative selection, pioneered by \citet{schwartzstein2021using}: an agent receiving a piece of information chooses how to interpret it, selecting the best-fitting information-generating process among the alternatives that could have generated it.\footnote{Differently from
\citet{schwartzstein2021using}, we do not consider a strategic Sender providing alternative
narratives, and we consider a specific class of information processes.} We ask which heuristic for selecting narratives yields the best evolutionary performance. We find that reconciling evidence with one's own prior beliefs is usually a poor rule, whereas treating the average belief in the population as an anchor is a considerably better one.

We also contribute to the literature on opinion dynamics, pioneered by
\citet{degroot1974reaching}, which studies how agents update their beliefs based on their peers'
opinions. For instance, \citet{forster2013anonymous}, \citet{buechel2015opinion}, and
\citet{grabisch2019model} examine convergence to a common belief in the presence of conformist
and anti-conformist agents. Our model incorporates conformism, assuming that conformist agents
treat the average belief in the population as an anchor. In contrast to the standard approach,
conformists' beliefs do not adjust directly toward the anchor: what the anchor governs is the narrative
they select, and hence how they interpret the information they receive.
The way conformism operates in our model also distinguishes it from herding in
financial markets. The canonical models of informational cascades show how social learning can break down once public signals outweigh private ones, so that choices cease to reveal private information \citep{bikhchandani1992theory,banerjee1992simple}. 
By contrast, in our model, conformist investors continue to update on their private information: the crowd affects how they interpret that information rather than whether they use it. 

We also contribute to the literature on the coexistence of different learning rules, showing that different heuristics for selecting narratives coexist in the long run. \citet{heller2015three}
shows that agents with different foresight abilities might coexist in a Prisoner's Dilemma with
an uncertain horizon.
Heterogeneous trading heuristics also coexist in financial markets
\citep{brock1997rational,chiarella2002heterogeneous,chiarella2003dynamics,anufriev2012evolution,anufriev2012evolutionary}. Coexistence matters for our mechanism, since the anchor is only useful if the population remains heterogeneous. \citet{dindo2020wisdom} face the same tension
in financial markets, where the wisdom of the crowd can arise only if selection
does not eliminate all but the most accurate agent. This tension
is the central question of the market selection hypothesis, which asks whether
competitive markets favor traders with more accurate beliefs. That literature typically finds that selection concentrates wealth on the single most accurate trader; however, this conclusion may fail with incomplete markets \citep{sandroni2000markets,sandroni2005efficient,blume2006youre}. We instead show that heterogeneous heuristics for
selecting narratives survive, and that anchoring one's narrative choice to the
population's average belief is favored by such competition precisely because
selection keeps that average belief close to the truth.

Finally, we contribute to the debate on confirmation bias, which may arise from the Bayesian
updating process \citep{wilson2014bounded,fryer2019updating} or in other forms
\citep{rabin1999first}. Relatedly, \citet{massari2026underreaction} shows that under-reaction may be
rewarded by financial markets under model misspecification rather than leading
behavioral agents to extinction. In our model, the advantage of under-confidence
depends on the underlying uncertainty: dampening information performs better
when moderate beliefs are more accurate. Under-confident agents outperform over-confident agents unless the state is highly predictable, whereas conformist agents outperform both by anchoring to the population's average belief.

\section{Model}\label{sec_mod}

We study a dynamic model with $n$ agents. We denote by $I\coloneqq\{1, \dots, n\}$ the set of agents. Time is discrete, and we index it with $t \in \mathbbm{N}$. 

\paragraph{State of the World, Beliefs and Information} The state of the world is the realization of a binary random variable: $\omega \in \Omega\coloneqq\{ A, B\}$. We denote with $\omega_t$ its realization at time $t$, and with $\hat{\omega}_t\coloneqq\{\omega_1,\omega_2,\dots,\omega_{t}\}$ the history of the state realizations up to time $t$. 
At time $t=0$, agents hold a common prior belief $\mu_0\in(0,1)$ that the state is $\omega=A$. At each time $t$, each agent $i\in I$ receives a signal $s_{i,t}\in S\coloneqq\{a,b\}$. Signals are independent and identically distributed across agents. Each signal originates from the same information-generating process $\pi$. In particular, $\pi$ is a function mapping states into distributions over messages. For any signal $s$ and any state $\omega$, $\pi(s|\omega)$ is the conditional probability of observing signal $s$ when the state is $\omega$. The information-generating process $\pi$ has an exogenous precision $p \in \left(\frac{1}{2},1\right)$, that is, the probability that the signal matches the state, i.e., $\pi(a|A)=\pi(b|B)=p$. 

We assume that some agents are subject to information misspecification: they may not know $p$ and think that the precision is either $\rho_1$ or $\rho_2$, such that $\frac{1}{2}<\rho_1<p<\rho_2<1$. Specifically, at each time $t$, these agents choose how to interpret the signal $s_{i,t}$, in other words, they choose a narrative $\rho_{i,t}\in\{\rho_1,\rho_2\}$, according to different heuristics. At each time $t$, all agents update their beliefs using Bayesian updating after receiving information. However, each agent may update their belief differently. Agent $i$ belief's updating depends on the signal $s_{i,t}$ they receive and the narrative $\rho_{i,t}$ they choose. We denote by $x_{i,t}\coloneqq(s_{i,t},\rho_{i,t})$ a vector indicating the information received by agent $i$ and their interpretation of the latter at time $t$, and by $h_{i,t} \coloneqq \left\{x_{i,1}, x_{i,2},\dots,x_{i,t-1} \right\}$ the history of agent $i$ at time $t$. 
Therefore, at each time $t$, the posterior belief of agent $i$ that $\omega=A$ is:
\begin{equation}
\label{Bayesian}
    \mu_{i,t}(s_{i,t},h_{i,t},\rho_{i,t})\coloneqq\frac{\pi_{\rho_{i,t}}(s_{i,t}|A)\mu_{i,t-1}(h_{i,t})}{\pi_{\rho_{i,t}}(s_{i,t}|A)\mu_{i,t-1}(h_{i,t})+\pi_{\rho_{i,t}}(s_{i,t}|B)(1-\mu_{i,t-1}(h_{i,t}))}
\end{equation}
We denote by $\pi_{\rho_{i,t}}$ the information-generating process associated with the narrative $\rho_{i,t}$. It holds that $\pi_{\rho_{i,t}}(a|A)=\pi_{\rho_{i,t}}(b|B)=\rho_{i,t}$ and $\mu_{i,-1}(h_{i,0})=\mu_0$ for any $i\in I$. 

\paragraph{Agents' Types.}
Agents are heterogeneous in their information processing methods. In particular, we distinguish five types:
\begin{enumerate}
    \item \textbf{Standard agents}. They know the correct information-generating process, which has precision $p$. Thus, it holds $\rho_{i,t}=p$ for any $t$.
    \item \textbf{Over-confident agents}. We follow \cite{schwartzstein2021using} and assume that, at each time $t$, each over-confident agent $i$ chooses their narrative $\rho_{i,t}$, after observing the signal's realization $s_{i,t}$, considering which narrative is more likely to have generated $s_{i,t}$.
The fit of a narrative at time $t$ given signal $s_{i,t}$ is defined as follows:
\begin{equation}
\label{fit}
    P(\rho_{i,t})\coloneqq\mu_{i,t-1}\pi_{\rho_{i,t}}(s_{i,t}|A)+(1-\mu_{i,t-1})\pi_{\rho_{i,t}}(s_{i,t}|B)
\end{equation}
where $\mu_{i,t-1}$ is agent $i$'s prior belief that $\omega=A$ when they choose the narrative. Thus, agent $i$ chooses $\rho_{i,t}$ that maximizes equation \eqref{fit}.\footnote{Equivalently, over-confident agents maximize the likelihood of the observed signal over a misspecified model, i.e., over the two narratives $\{\rho_1,\rho_2\}$, neither of which is the true precision $p$.}
\item \textbf{Under-confident agents}. At each time $t$, each under-confident agent $i$ chooses their narrative $\rho_{i,t}$, after observing the signal's realization $s_{i,t}$, considering which narrative is less likely to have generated $s_{i,t}$. Thus, agent $i$ chooses $\rho_{i,t}$ that minimizes equation \eqref{fit}.
\item \textbf{Conformist agents}. At each time $t$, each conformist agent $i$ chooses their narrative $\rho_{i,t}$, after observing the signal's realization $s_{i,t}$, to match their posterior belief with $\bar{\mu}_t$, the average posterior belief in the population at time $t$. We assume that conformist agents are myopic, meaning that, at each time $t$, their expectation of the average posterior belief coincides with the average prior belief, that is $E[\bar{\mu}_t]=\bar{\mu}_{t-1}(h_t)$, which each agent observes. We denote by $h_t \coloneqq \left\{h_{1,t}, h_{2,t}, \dots, h_{n,t} \right\}$ the vector of all agents' histories up to $t$. The utility of a conformist agent $i$ at time $t$ is:
\begin{equation}
\label{u_conf}
    C_{i,t}(\rho_{i,t},s_{i,t},h_{t})=-(\mu_{i,t}(s_{i,t},h_{i,t},\rho_{i,t})-\bar{\mu}_{t-1}(h_t))^2
\end{equation}
Thus, agent $i$ chooses $\rho_{i,t}$ that maximizes equation \eqref{u_conf}.
\item \textbf{Anti-conformist agents}. At each time $t$, each anti-conformist agent $i$ chooses their narrative $\rho_{i,t}$, after observing the signal's realization $s_{i,t}$, to maximize the distance between their posterior belief and $\bar{\mu}_t$. Similarly to conformist agents, anti-conformist agents are also myopic and, thus, $E[\bar{\mu}_t]=\bar{\mu}_{t-1}(h_t)$. It follows that the utility of an anti-conformist agent $i$ at time $t$ is $-C_{i,t}(\rho_{i,t},s_{i,t},h_{t})$. Thus, agent $i$ chooses $\rho_{i,t}$ that minimizes equation \eqref{u_conf}.
\end{enumerate}

\paragraph{Evolution of the State.} A parameter $q\in[0,1]$ governs the evolution of the state. We define a law of motion---henceforth referred to as LoM---$f(\omega_t;\hat{\omega}_{t-1}, h_t)$ as follows: for any time $t$, $f(\omega_t;\hat{\omega}_{t-1}, h_t)$ is the probability that $\omega_t$ is the realization of the state at time $t$ given the realizations of the state up to time $t-1$ and the histories of all agents at time $t$.
In the following, we consider three LoMs: 
\begin{enumerate}
    \item \textbf{Independent State:} $f(A;\hat{\omega}_{t-1}, h_t)=f(A)=q$. The state realization is a random draw in each period. Signals are independent across time. 
    \item \textbf{Auto-Correlated State:}
$f(\omega_t;\hat{\omega}_{t-1}, h_t)=f(\omega_t|\omega_{t-1})$ and $f(\omega_t=k|\omega_{t-1}=k)=\phi_k$ for $k=A,B$ and $q$ represents the invariant probability that $\omega=A$.\footnote{This corresponds to the following transition matrix:
\begin{equation*}
\label{transition}
P=\left(
\begin{array}{cc}
\phi_A   & 1-\phi_A \\
1-\phi_B  & \phi_B
\end{array}
\right)
\end{equation*}
where each row represents the state at time $t-1$, whereas each column represents the state at time $t$.
Define $\nu=\left(\nu_1;\nu_2\right)$ such that:
\begin{equation*}
    \nu=\nu P, \quad \quad \nu_1=\frac{1-\phi_B}{2-\phi_A-\phi_B}=q, \quad \quad
    \nu_2=\frac{1-\phi_A}{2-\phi_A-\phi_B}.
\end{equation*}
We normalize $\phi_B=\frac{1}{2}$, so that the evolution of the state only
depends on $q$. We define $\phi_A$ as a function of $q$:
\begin{equation*}
    q=\frac{1}{3-2\phi_A} \iff \phi_A=\frac{3q-1}{2q}
\end{equation*}
} The state realization at time $t$ depends on the state realization at time $t-1$. Unlike under the LoM \say{Independent State}, where the realization does not depend on the past, the state is thus partly predictable from its own history.
Signals are independent across time.
\item \textbf{Self-fulfilling State:} $f(A;\hat{\omega}_{t-1}, h_t)=f(A|h_t)=\delta \bar{\mu}_{t-1}(h_t) + (1-\delta) q$, where $\delta \in [0,1]$. The state realization depends on $q$ as well as on the average prior belief in the population, that is, $\bar{\mu}_{t-1}(h_t)$. $\delta$ represents the weight of $\bar{\mu}_{t-1}(h_t)$ in determining the state's realization. In other words, $\delta$ is the relative degree of endogeneity of the state realization with respect to agents' prior beliefs---that is, it represents to what extent the state is a ``self-fulfilling prophecy''. It follows that signals are dependent across time through $\bar{\mu}_{t-1}(h_t)$.
\end{enumerate}

\paragraph{Evolution of the Population.}
We assume that the initial population is a random partition into the five agent types. 
We also assume that the composition of the population changes over time, particularly every $\tau$ periods; we refer to the times $t=\tau, 2\tau, \dots$ when this occurs as turnover times. An agent's survival depends on their performance relative to the population average. We define $\epsilon_{i,t}$ as the squared error of agent $i$ at time $t$:
\begin{equation}
\label{evolutionary_fitness}
    \epsilon_{i,t}\coloneqq (q-\mu_{i,t})^2
\end{equation} 
and $\psi_t = \frac{1}{n}\sum_{i=1}^n \epsilon_{i,t}$ as the mean squared error in the population.
An agent $i$ survives at each turnover time with probability $1-g(\epsilon_{i,t}; \psi_t)$, where:
\begin{equation}
\label{survival}
        g(\epsilon_{i,t}; \psi_t)\coloneqq\begin{cases}
        \dfrac{1}{2}\,\dfrac{\epsilon_{i,t} - \underline{\epsilon}}{\psi_t - \underline{\epsilon}}, & \underline{\epsilon} \leq \epsilon_{i,t} \leq \psi_t, \\[8pt]
        \dfrac{1}{2} \left(1+\dfrac{\epsilon_{i,t}-\psi_t}{\bar{\epsilon}-\psi_t}\right), & \psi_t \leq \epsilon_{i,t} \leq \bar{\epsilon}.
        \end{cases}
    \end{equation}
    where $\underline{\epsilon}=\min\{\epsilon_{i,t}\}_{i\in I}$, and $\bar{\epsilon}=\max\{\epsilon_{i,t}\}_{i\in I}$. Thus, the probability of survival is a continuous and decreasing function of the agent's error. This assumption reflects the idea that agents who make bigger mistakes are more likely to exit the market.
    Our survival selection criterion is coherent with a \say{survival of the fittest} interpretation, which is standard in evolutionary studies.
    We also assume that the total population, $n$, remains constant over time. Therefore, we substitute each agent who exits the market with a newborn agent having a random type and a prior belief $\mu_0$. We denote by $\varphi_t\in(0,1)$ the realized share of the population replaced at turnover time $t$.

\paragraph{System Neglect.} Survival depends on how precisely an agent predicts the likelihood of each state, that is, on their error with respect to $q$. Agents, however, form beliefs about the realization $\omega_t$,  not about the probability $q$ governing it: they are unaware of $q$ and observe only evidence about the current realization. This is an extreme form of system neglect \citep{massey2005detecting}: agents react to the signals they observe, whereas the process generating those signals never enters their belief space. This is coherent with \citet{griffin1992weighing}, who show that judgments are driven by the strength of evidence with
comparatively little consideration given to its statistical weight. \citet{kremer2011demand} document similar behavior in a forecasting task where, as here, the parameters of the process are not communicated to subjects.  Agents suffer no misspecification when $q=1$, since beliefs about $\omega_t$ and about $q$ coincide; we exploit this case in Section~\ref{sec_bench_naive/conf}.

\subsection{Optimal Narrative}\label{subsec_optnarr}
For those agents who do not know $p$, we study their problem of identifying the optimal narrative, that is, the optimal $\rho_{i,t}$. 

\paragraph{Over-confident and Under-confident agents.} At each time $t$, the fit of a narrative---defined by equation \eqref{fit}---is a linear function of an agent $i$'s prior belief $\mu_{i,t-1}$. Therefore, an over-confident agent maximizes the fit for a given signal $s_{i,t}$ if they choose $\rho_{i,t}$ such that they maximize the probability $\pi_{\rho_{i,t}}(s_{i,t}|\omega)$ under the state realization $\omega$ that they consider most likely given their prior belief. Formally,
\begin{equation*}
    \rho_{i,t}=\left\{\begin{array}{cc}
      \rho_1   &  \mbox{if } (\mu_{i,t-1}\geq \frac{1}{2} \land s_{i,t}=b) \lor (\mu_{i,t-1}\leq \frac{1}{2} \land s_{i,t}=a)\\
      \rho_2   &  \mbox{if } (\mu_{i,t-1}\geq \frac{1}{2} \land s_{i,t}=a) \lor (\mu_{i,t-1}\leq \frac{1}{2} \land s_{i,t}=b)
    \end{array}\right.
\end{equation*}
We define a signal $s_{i,t}$ as ``pro-attitudinal'' (resp., ``counter-attitudinal'') if it
reinforces (resp., weakens) an agent's prior belief. We say that an agent amplifies a
signal when they attribute to it the high precision $\rho_2$, and dampens it when they
attribute to it the low precision $\rho_1$. An over-confident agent thus amplifies
pro-attitudinal signals and dampens counter-attitudinal ones. Under-confident agents choose the
narrative that minimizes the fit, and therefore do the opposite.

\paragraph{Conformist and Anti-conformist agents.}
Each conformist chooses the narrative $\rho_{i,t}$ that minimizes the distance between their posterior belief $\mu_{i,t}(s_{i,t},h_{i,t},\rho_{i,t})$ and the average prior belief in the population, $\bar{\mu}_{t-1}(h_t)$. The optimal narrative depends on the signal received, and we characterize it separately for pro-attitudinal and counter-attitudinal signals. Without loss of generality, let $\mu_{i,t-1}\geq\frac{1}{2}$, so that signal $a$ is pro-attitudinal and signal $b$ counter-attitudinal (the case $\mu_{i,t-1}\leq\frac{1}{2}$ is symmetric). By equation~\eqref{Bayesian}, both narratives move the posterior above the prior after the pro-attitudinal signal and below it after the counter-attitudinal one, with $\rho_2$ producing the larger movement in either case. 
We denote by $\Xi_{i,t}(s_{i,t})$ the interval with endpoints $\mu_{i,t}(s_{i,t},h_{i,t},\rho_1)$ and $\mu_{i,t}(s_{i,t},h_{i,t},\rho_2)$, that is, the set of posteriors attainable after the realized signal, and by $\overline{M}(\mu_{i,t-1})$ and $\underline{M}(\mu_{i,t-1})$ its midpoint after a pro-attitudinal and a counter-attitudinal signal, respectively. Remarkably, the length of $\Xi_{i,t}(s_{i,t})$ is increasing in $\rho_2$, decreasing in $\rho_1$, and vanishes as $\rho_2\to\rho_1$. Since
$\rho_1<p<\rho_2$, it holds that $\mu_{i,t}(s_{i,t},h_{i,t},p)\in\Xi_{i,t}(s_{i,t})$. The following lemma characterizes the optimal narrative of conformist agents.

\begin{lemma}\label{lem_constrained}
Consider a conformist agent with prior belief $\mu_{i,t-1}\geq\frac{1}{2}$ and realized signal $s_{i,t}$.
\begin{enumerate}
    \item[(i)] If $s_{i,t}=a$, the agent chooses $\rho_1$ if $\bar{\mu}_{t-1}(h_t)<\overline{M}(\mu_{i,t-1})$ and $\rho_2$ if $\bar{\mu}_{t-1}(h_t)>\overline{M}(\mu_{i,t-1})$. 
    \item[(ii)] If $s_{i,t}=b$, the agent chooses $\rho_1$ if $\bar{\mu}_{t-1}(h_t)>\underline{M}(\mu_{i,t-1})$ and $\rho_2$ if $\bar{\mu}_{t-1}(h_t)<\underline{M}(\mu_{i,t-1})$. 
    \item[(iii)] 
    If $\bar{\mu}_{t-1}(h_t)\notin\Xi_{i,t}(s_{i,t})$, then $|\mu_{i,t}(s_{i,t},h_{i,t},\rho_{i,t})-\bar{\mu}_{t-1}(h_t)|<|\mu_{i,t}(s_{i,t},h_{i,t},p)-\bar{\mu}_{t-1}(h_t)|$.
\end{enumerate}
\end{lemma}

A conformist agent amplifies or dampens a signal depending on which narrative brings their
belief more in line with the average prior belief in the population $\bar{\mu}_{t-1}(h_t)$. Lemma \ref{lem_constrained} (i) and (ii) show that a conformist agent chooses amplification if and only if $\bar{\mu}_{t-1}(h_t)$ lies beyond (below) the midpoint of the two attainable posteriors when the signal is pro-attitudinal (counter-attitudinal), and dampens otherwise. Intuitively, the two narratives are equally attractive at the midpoint; thus, if the signal is pro-attitudinal (counter-attitudinal), amplification pays when the conformist agent's prior is sufficiently less (more) extreme than the population's average belief.

We can interpret conformism as an endogenous switching between different heuristics. A conformist behaves like an under-confident agent if and only if $\bar{\mu}_{t-1}(h_t)<\underline{M}(\mu_{i,t-1})$, that is, when the conformist is far more extreme than the crowd, and like an over-confident agent if and only if $\bar{\mu}_{t-1}(h_t)>\overline{M}(\mu_{i,t-1})$, that is, when the conformist is far more moderate. Instead, a conformist near the crowd, with $\bar{\mu}_{t-1}(h_t)\in\big(\underline{M}(\mu_{i,t-1}),\overline{M}(\mu_{i,t-1})\big)$, always selects $\rho_1$, dampening all signals to stay close to the crowd. 

Lemma \ref{lem_constrained} (iii) shows that, whenever the average belief in the population lies outside the set of posteriors a conformist can attain, the latter ends up closer to the crowd than a standard agent facing the same prior and signal. Narrative selection thus grants conformist agents an advantage in pursuing their objective, but only to the extent that the two narratives permit: the agent can only reach the endpoints of $\Xi_{i,t}(s_{i,t})$, so the average belief remains out of reach. 
When instead $\bar{\mu}_{t-1}(h_t)\in\Xi_{i,t}(s_{i,t})$, a conformist agent may overshoot or undershoot the anchor, and end up further from it than a standard agent.

Anti-conformist agents choose the narrative that maximizes the distance between their posterior belief and $\bar{\mu}_{t-1}(h_t)$; their choice is the pointwise mirror of conformist agents' after each signal, so that for $\bar{\mu}_{t-1}(h_t)\in\big(\underline{M}(\mu_{i,t-1}),\overline{M}(\mu_{i,t-1})\big)$ they choose $\rho_2$ after either signal, whereas they follow the over-confident heuristic for $\bar{\mu}_{t-1}(h_t)<\underline{M}(\mu_{i,t-1})$ and the under-confident one for $\bar{\mu}_{t-1}(h_t)>\overline{M}(\mu_{i,t-1})$.

\subsection{Application: Financial Markets}\label{subsec_laws}

Financial markets provide a natural setting for our model. A pool of investors must decide
whether to purchase a firm's stock, whose price (and hence each investor's profit) depends on
how the firm performs. Each investor receives imperfectly informative news
about the firm's performance (e.g., media coverage, direct analysis of the firm's
reports, market information such as stock prices), and different investors need not see the same news. Investors must decide how much to trust that news, and this is what the choice of a narrative captures: $\rho_1=\mbox{``This news is just noise''}$ and $\rho_2=\mbox{``This news is about the firm's fundamentals''}$.

Investors use different heuristics. An over-confident investor trusts the news that supports their prior view of the firm and dismisses the rest. An under-confident investor does the opposite, wary of their own thesis and taking contrary evidence seriously. Conformist and anti-conformist investors instead judge news by whether their view of the firm is more or less extreme than the market's, for which market prices and consensus forecasts provide observable proxies.

Investors' success depends on their ability to predict how likely the firm is to perform well,
that is, $q$, rather than on calling any single realization of its performance. An investor who overestimates $q$ takes a larger position in the stock than its prospects warrant, and one who underestimates it forgoes a profitable position; either way, persistent misjudgment of $q$ erodes returns relative to peers, and the investor is more likely to exit the market. Yet investors assess the firm's current performance, not its underlying probability of success; that is, they suffer from system neglect. Consistent with this, \citet{greenwood2014expectations} find that investors' expectations of future stock market returns track recently realized returns rather than
model-based expected returns. More broadly, under- and overreaction to information are longstanding themes in
finance \citep{debondt1985does,daniel1998investor}. The process neglected by investors can take different forms: the firm's performance may be independent across periods, path-dependent, or partly driven by investors' own expectations (since the stock price reflects not only the firm's profitability but also the market's assessment of it).

\section{Theoretical Benchmark}
\label{sec_bench_naive/conf}

In this section, we analytically characterize the mechanisms underlying our results. Conformism
pulls beliefs toward a population anchor, and whether it improves evolutionary performance
depends on how close the anchor is to $q$. To isolate each ingredient of the full model, we explore three configurations that differ only in the anchor available to conformist agents and in whether they reach it through narrative selection:
\begin{itemize}
    \item[\textbf{A.}] \emph{Rational-expectations anchor (Section~\ref{subsec_configA}).} Conformist agents anchor mechanically to the contemporaneous expected average belief, $E\big[\bar{\mu}_{t}(h_{t+1})\mid h_t\big]$.
    \item[\textbf{B.}] \emph{Adaptive anchor (Section~\ref{subsec_configB}).} Conformist agents anchor mechanically to the lagged realized average belief, $\bar{\mu}_{t-1}(h_t)$, as in the main model.
    \item[\textbf{C.}] \emph{Adaptive anchor with narrative selection (Section~\ref{subsec_configC}).} Conformist agents can no longer anchor mechanically, but rather through the choice of a narrative $\rho_{i,t}\in\{\rho_1,\rho_2\}$, as characterized by Lemma~\ref{lem_constrained}.
\end{itemize}
In Sections~\ref{subsec_configA}--\ref{subsec_configC}, we maintain two assumptions. First, the population consists only of standard and conformist agents. Second, we consider the LoM \say{Independent State} with $q=1$, so that the state equals $A$ in each period and agents do not suffer from misspecification when updating their beliefs.\footnote{With $q=1$, the LoM \say{Auto-correlated State} coincides with
\say{Independent State}. The LoM \say{Self-fulfilling State} does not: the
probability of state $A$ remains below one whenever $\delta>0$ and
$\bar{\mu}_{t-1}(h_t)<1$.} Section~\ref{subsec_interior} considers $q<1$, and Section~\ref{subsec_heuristics} characterizes the remaining types.

A general trade-off underlies all three configurations.
By equation \eqref{evolutionary_fitness}, the expected error of agent $i$ at time $t$ decomposes as
\begin{equation}\label{eq_biasvariance}
    E\big[\epsilon_{i,t}\mid h_t\big]\;=\;\Big(E\big[\mu_{i,t}\mid h_t\big]-q\Big)^2+\mathrm{Var}\big(\mu_{i,t}\mid h_t\big),
\end{equation}
that is, in terms of bias and variance. Evolutionary selection rewards calibrated (low bias with respect to $q$) and stable (low dispersion across histories) beliefs.

Consider a conformist agent $i$ with prior belief $\mu^C_{i,t-1}$. We denote by $\tilde{\mu}_{i,t}$ their Bayesian belief, that is, the posterior belief they would hold if they updated according to equation~\eqref{Bayesian} with $\rho_{i,t}=p$, without conforming to the population, and by $\mu^C_{i,t}$ the posterior belief they actually hold, which in configurations A and B takes the form
\begin{equation}\label{eq_shrinkage}
    \mu^C_{i,t} \;=\; \lambda\, \tilde{\mu}_{i,t} \;+\; (1-\lambda)\, m_t, \qquad \lambda\in(0,1),
\end{equation}
where $m_t$ denotes the anchor: $m_t=E\big[\bar{\mu}_{t}(h_{t+1})\mid h_t\big]$ in configuration A and $m_t=\bar{\mu}_{t-1}(h_t)$ in configuration B. 
Finally, we write $b_{i,t}\coloneqq E[\tilde{\mu}_{i,t}\mid h_t]-q$ for the bias of agent $i$'s Bayesian belief, $\sigma^2_{i,t}\coloneqq\mathrm{Var}(\tilde{\mu}_{i,t}\mid h_t)>0$ for its variance, and $d_t\coloneqq m_t-q$ for the bias of the anchor.
The following proposition characterizes the instantaneous effect of conformism. 

\begin{proposition}\label{prop_shrinkage}
Let $\Delta_{i,t} \coloneqq E[(\mu^C_{i,t}-q)^2\mid h_t]-E[(\tilde{\mu}_{i,t}-q)^2\mid h_t]$ denote the change in expected error caused by conforming. Then
\begin{equation}\label{eq_delta}
    \Delta_{i,t} \;=\; \big(\lambda b_{i,t} + (1-\lambda) d_t\big)^2 - b_{i,t}^2 - (1-\lambda^2)\,\sigma^2_{i,t}  ,
\end{equation}
and:
\begin{enumerate}
    \item[(i)] if $|d_t| \leq |b_{i,t}|$, then $\Delta_{i,t} < 0$;
    \item[(ii)] if $|d_t| > |b_{i,t}|$, then $\Delta_{i,t} < 0$ if and only if
    $\sigma^2_{i,t} > \bar{\sigma}^2_{i,t} \coloneqq \frac{(\lambda b_{i,t} + (1-\lambda) d_t)^2 - b_{i,t}^2}{1-\lambda^2}$.
\end{enumerate}
\end{proposition}

Conformism always reduces belief variance, whereas its effect on bias depends on the anchor. If the anchor is at least as calibrated as the agent's Bayesian belief, conformism unambiguously improves evolutionary performance. If the anchor is less calibrated, conformism trades a deterioration in bias for a reduction in variance, and it improves evolutionary performance only when the agent's Bayesian belief is sufficiently dispersed. 

Conditioning on $h_t$ makes the anchor a constant. The comparison between $|d_t|$ and $|b_{i,t}|$ locates the agent relative to the crowd, given the history that has produced their belief. Averaging over histories, the anchor becomes a random variable, and the variance in equation \eqref{eq_biasvariance} has two additional terms: $(1-\lambda)^2\mathrm{Var}(m_t)+2\lambda(1-\lambda)\mathrm{Cov}\big(\tilde{\mu}_{i,t},m_t\big)$. When $q=1$, the state is degenerate, so signals are independent across agents, and the average belief aggregates $n$ idiosyncratic histories, making both terms of order $1/n$. Proposition~\ref{prop_shrinkage} therefore describes average errors as well, up to a term of that order. The same is not true when $q<1$: the state is now random, so signals are correlated across agents, and the variance of the average belief does not vanish as $n$ grows; Section~\ref{subsec_interior} returns to this point. 

Proposition~\ref{prop_shrinkage} concerns a conformist agent at a given time $t$, whereas the next results will compare the evolutionary performance of a conformist agent and a standard agent.

\subsection{Rational-Expectations Anchor}\label{subsec_configA}

We first endow conformist agents with the best conceivable anchor, that is, $m_t=E\big[\bar{\mu}_{t}(h_{t+1})\mid h_t\big]$.
Since $q=1$, the error $\epsilon_{i,t}=(1-\mu_{i,t})^2$ is decreasing in the belief: holding a
belief closer to $1$ is unambiguously better. The anchor conformist agents rely on is bounded
away from certainty: at each turnover time $t$, a share $\varphi_t\geq\frac{1}{n}$ of the
population consists of newborn agents holding the prior belief $\mu_0<1$. Hence,
\begin{equation}\label{eq_anchorbound}
    \bar{\mu}_{t}(h_{t+1})\;\leq\;1-\varphi_t(1-\mu_0)\;\leq\;1-\frac{1-\mu_0}{n}\;\eqqcolon\;\kappa\;<\;1.
\end{equation}
The bound $\kappa$ exists for any prior $\mu_0<1$.
If newborn agents adopt the average belief, as in Section~\ref{subsec_rob}, equation \eqref{eq_anchorbound} becomes $\bar{\mu}_{t}(h_{t+1})\leq1-\varphi_t\big(1-\bar{\mu}_{t-1}(h_t)\big)$: newborn agents no longer pull the average belief back toward $\mu_0$, but only slow its rise, so iterating over $k$ turnover times yields $\bar{\mu}_{t}(h_{t+1})\leq1-n^{-k}(1-\mu_0)$, which converges to $1$ as $k\to\infty$.

\begin{proposition}\label{prop_configA}
Let $q=1$ and consider a conformist agent $i$ and a standard agent $j$, with $\mu^C_{i,0}=\mu^S_{j,0}$ and $h_{i,t}=h_{j,t}$ at every time $t$. Then:
\begin{enumerate}
    \item[(i)] $\epsilon_{i,t}\geq(1-\lambda)^2(1-\kappa)^2>0$ infinitely often, so that
    $\epsilon_{i,t}\nrightarrow0$, whereas $\epsilon_{j,t}\rightarrow0$ almost surely;
    \item[(ii)] If $m_t\leq\mu^S_{j,t}$ at every $t$, then $\mu^C_{i,t}\leq\mu^S_{j,t}$ and hence $\epsilon_{i,t}\geq\epsilon_{j,t}$, at every $t$.
\end{enumerate}
\end{proposition}

Proposition \ref{prop_configA} (i) shows that conformism is a disadvantage when $q=1$ because a conformist agent's belief is tied to an anchor that turnover keeps away from certainty. Therefore, their error does not vanish, whereas a standard agent's does. Proposition \ref{prop_configA} (ii) sharpens the comparison, ranking the two agents at every time whenever the average belief lies below the standard agent's.

When Proposition~\ref{prop_configA} (ii) fails, the ranking is reversed, and conformism becomes an
advantage. This happens whenever the standard agent's belief lies below the average, for two
reasons. The first is age: newborn agents hold the prior belief $\mu_0$, whereas incumbents have
(on average) updated upward from it. The second is luck: an agent whose history has been
unfavorable lies below the average regardless of their age. In both cases, anchoring pulls the
belief up toward the crowd and protects the conformist agent from the resulting error.

This bears on the coexistence of types. Because agents are replaced at each turnover time, the population always contains recent entrants, who lie below the crowd. Therefore, the condition under which conformism pays keeps recurring. Selection reinforces the point: an unlucky standard agent incurs a large error and is likely to exit the market, whereas an unlucky conformist agent, being anchored, incurs a smaller error and is more likely to survive. Hence, conformist agents persist when $q=1$ not because they eventually catch up, which Proposition~\ref{prop_configA} (i) rules out, but because the population always contains agents whose beliefs are less calibrated than the crowd's.

\subsection{The Adaptive Anchor}\label{subsec_configB}

In the main model, the rational-expectations anchor is unavailable to conformist agents, who observe the average belief from the previous period rather than form expectations for the current period. Let $m_t=\bar{\mu}_{t-1}(h_t)$. Proposition \ref{prop_configA} continues to hold: its proof uses only that equation~\eqref{Bayesian} is increasing in the prior belief and that the anchor is bounded above, by $\kappa$ in part (i) and by the standard agent's belief in part (ii). Both properties hold path by path for $\bar{\mu}_{t-1}(h_t)$. Thus, whether a conformist agent correctly anticipates the crowd or observes it with a delay does not determine the comparison with a standard agent when $q=1$, which licenses us to use the adaptive anchor in our simulations.

\subsection{Narrative Selection}\label{subsec_configC}

In the main model, conformist agents cannot implement equation~\eqref{eq_shrinkage}: they can
only select the narrative that brings their posterior the closest to $\bar{\mu}_{t-1}(h_t)$.
Therefore, the anchor no longer bounds a conformist agent's belief. Note that, by equation~\eqref{Bayesian}, the log-odds of a belief, $\ln\frac{\mu_{i,t}}{1-\mu_{i,t}}$, move by $\pm\ln\frac{\rho_{i,t}}{1-\rho_{i,t}}$ each period, independently of the current belief.

The comparison with a standard agent turns on the precision that the conformist agent attributes to their signal, which Lemma~\ref{lem_constrained} ties to their position relative to the crowd. An agent far more moderate than the crowd follows the over-confident heuristic and thus converges
to $q$ faster than a standard agent, since $\rho_1<p<\rho_2$; as they approach the crowd, they
begin to dampen all signals, and their belief still rises, only more slowly. Turnover keeps the anchor away from certainty, so they tend to end up more extreme than the crowd, and from then on they follow the under-confident heuristic, dampening pro-attitudinal signals and amplifying counter-attitudinal ones. In this last stage, the agent's expected log-odds drift is:
\begin{equation}\label{eq_drift}
    p\,\ln\frac{\rho_1}{1-\rho_1}-(1-p)\,\ln\frac{\rho_2}{1-\rho_2},
\end{equation}
which may be negative: a conformist agent behaving as an under-confident agent can be pushed away from certainty rather than merely approach it slowly. In that case, their belief is pulled back toward the crowd and cannot durably run ahead of it. Standard agents are favored either way: the disadvantage from conformism vanishes when the drift in equation \eqref{eq_drift} is positive, but persists when it is negative.

\subsection{Uncertainty}\label{subsec_interior}

We now relax the assumption of no uncertainty and let $q<1$. Thus, the error $\epsilon_{i,t}=(q-\mu_{i,t})^2$ is no longer decreasing in the belief; a standard agent's belief no longer converges to $q$.

\begin{proposition}\label{prop_interior}
Let $q\in(\frac{1}{2},1)$. For any standard agent $j$, $\mu^S_{j,t}\to1$ almost surely; hence, almost surely, $b_{j,t}\to1-q>0$, $\sigma^2_{j,t}\to0$, and $E[\epsilon_{j,t}\mid h_t]\to(1-q)^2$.
\end{proposition}

A standard agent accumulates evidence in favor of the more frequent state, and their belief converges to certainty. Evolutionary performance, however, rewards calibration to the state's probability, so this convergence is costly. A conformist agent, anchored to the average belief, does not converge to certainty, and this now works to their advantage.

By Proposition~\ref{prop_shrinkage}, conformism strictly reduces the expected error whenever
$|d_t|\leq|b_{i,t}|$, that is, whenever the anchor $m_t$ is at least as close to $q$ as the
Bayesian belief $\tilde{\mu}_{i,t}$. The latter converges to certainty under any $q$, whereas the
anchor remains in the interior. When $q=1$, certainty is the accurate belief and conformism raises
the expected error; instead, when $q<1$, conformism lowers it. Neither conclusion depends on
whether $m_t=E\big[\bar{\mu}_t(h_{t+1})\mid h_t\big]$ or $m_t=\bar{\mu}_{t-1}(h_t)$.

By Lemma~\ref{lem_constrained}(iii), whenever the anchor lies outside the set of attainable
posteriors, conforming selects the one closest to it. Thus, the advantage of conformism 
survives the restriction to two narratives if the anchor is close to $q$. By Proposition~\ref{prop_interior}, standard agents would carry the anchor to certainty; the following observation identifies the force that keeps it near $q$.

\begin{remark}\label{rem_calibrationnew}
Evolutionary selection recurrently concentrates beliefs around $q$: at any turnover time, the error $\epsilon_{i,t}$ among survivors is first-order stochastically dominated by its pre-selection counterpart.
\end{remark}

Remark~\ref{rem_calibrationnew} states that conformism is not advantageous because crowds are exogenously wise, but because the same evolutionary process that selects agents also disciplines the anchor. The anchor is shaped by two forces: selection, which favors agents whose beliefs are close to $q$, and turnover, which pulls the average belief back toward $\mu_0$. In Section \ref{subsec_implications}, P2 tests whether the anchor stays calibrated, and P5 isolates the first force by letting newborn agents adopt the average belief. P3 tests whether higher signal precision increases the anchor's volatility and thereby weakens the conformist advantage, particularly when $q$ is high and the more extreme beliefs of the other types are less costly.

As noted after Proposition~\ref{prop_shrinkage}, our results are conditional on the history
$h_t$, and the two additional variance terms that arise when averaging over histories vanish as
$n$ grows only when $q=1$. When $q<1$, Proposition~\ref{prop_shrinkage} only ranks the two agents at a given history, and the ranking of average errors is established in Section~\ref{sec_sim}.

\subsection{Over- and Under-confidence}\label{subsec_heuristics}

As shown in Section~\ref{subsec_optnarr}, an over-confident agent selects $\rho_2$ after pro-attitudinal signals and $\rho_1$ after counter-attitudinal ones; an under-confident agent does the opposite. Writing $L_1\coloneqq\ln\frac{\rho_1}{1-\rho_1}$ and $L_2\coloneqq\ln\frac{\rho_2}{1-\rho_2}$, the log-odds of an over-confident agent whose prior belief lies above $\frac12$ change by $+L_2$ after signal $a$ and $-L_1$ after signal $b$. Under-confident agents behaves oppositely, those whose prior belief lies above $\frac12$ change by $+L_1$ after signal $a$ and $-L_2$ after signal $b$.  Let $P_a\coloneqq qp+(1-q)(1-p)$ be the probability that a signal equals $a$. An over-confident agent's expected change in log-odds is $\bar{D}\coloneqq P_aL_2-(1-P_a)L_1$, and an under-confident agent's is $D\coloneqq P_aL_1-(1-P_a)L_2$, of which equation~\eqref{eq_drift} is the case $q=1$. When the prior belief is below $\frac12$, signal $b$ becomes the pro-attitudinal one, so an over-confident agent's expected log-odds change is $D$ and an under-confident agent's is $\bar{D}$.

\begin{proposition}\label{prop_heuristics}
Let $q\in(\frac{1}{2},1]$ and consider an agent $i$ with prior belief $\mu_{i,t-1}$. Then, whatever the side of $\frac{1}{2}$ on which $\mu_{i,t-1}$ lies:
\begin{enumerate}
    \item[(i)] if $D>0$, both an over-confident and an under-confident agent's belief is
    expected to move toward $1$;
    \item[(ii)] if $D<0$, an over-confident agent's belief is expected to become more
    extreme, and an under-confident agent's less extreme.
\end{enumerate}
Moreover, $D$ is increasing in $q$ and in $p$, and decreasing in $\rho_2-\rho_1$.
\end{proposition}

The sign of $D$ separates two regimes. When $D>0$, both heuristics carry beliefs toward $1$. When $D<0$, they pull apart: over-confident agents are driven outward on both sides, so an agent whose belief lies below $\frac{1}{2}$ becomes certain that the state is $B$ even though $A$ is more likely. Thus, their beliefs end up extreme on either side. Under-confident agents face the opposite force and remain moderate. Since $D$ is decreasing in $\rho_2-\rho_1$, this is the regime of high narrative heterogeneity: when the two available interpretations differ sharply, dampening pro-attitudinal signals makes a belief less extreme rather than merely slowing its rise.

The two regimes have different consequences for survival. When $D<0$, an under-confident agent's
belief stays near $\frac{1}{2}$, so their error is approximately $\left(q-\frac{1}{2}\right)^2$,
whereas an agent whose belief converges to certainty incurs an error approaching $(1-q)^2$: the
former is smaller if and only if $q<\frac{3}{4}$. Being insensitive to information is thus an
advantage when the state is close to unpredictable and a growing disadvantage as it becomes
predictable. Since $D$ is increasing in $q$, however, the force holding under-confident beliefs
near $\frac{1}{2}$ is weaker at high $q$. This works in their favor, since their beliefs become
less moderate exactly when moderation is costly, so they need not disappear as the state becomes
predictable.

\subsection{Predictions}\label{subsec_implications}

Below, we state predictions about evolutionary performance when $q<1$, which we test with simulations in Section~\ref{sec_sim}.

\begin{enumerate}
    \item[\textbf{P1.}] Conformist agents outperform standard agents. Proposition~\ref{prop_shrinkage} establishes this at a given history; the prediction extends it across histories.
    \item[\textbf{P2.}] Conformist agents dominate because the anchor remains close to
    $q$.
    \item[\textbf{P3.}] Raising the signal precision $p$ should reduce the
    conformist advantage if $q$ is sufficiently high.
    \item[\textbf{P4.}] Reducing narrative heterogeneity shrinks conformist agents' adjustment to the anchor (Lemma~\ref{lem_constrained}) and raises $D$ (Proposition~\ref{prop_heuristics}). Thus, as $\rho_2-\rho_1$ vanishes, conformist agents perform worse, whereas under-confident agents perform better at high $q$.
    \item[\textbf{P5.}] Conformist agents remain dominant even if newborn agents adopt the average belief rather than $\mu_0$, so that selection is the only force left holding the anchor near $q$.
    \item[\textbf{P6.}] Conformist agents outperform all other types, because the other heuristics hold the average belief away from the boundaries: under-confident agents remain moderate and over-confident agents are dispersed on both sides (Proposition~\ref{prop_heuristics}).
\end{enumerate}

\section{Simulations with Narrative Selection}\label{sec_sim}

In this section, we study the model using simulations, testing the predictions of Section~\ref{sec_bench_naive/conf} and extending the analysis to the five types. 

\paragraph{Procedures.} Table~\ref{tab_bench} reports the parameters used in the simulations:
an intermediate information precision $p$, two narratives with low and high precision, a
neutral prior belief $\mu_0$, and a stopping time $T$ large enough for the population to
stabilize.

\begin{table}[h!]
\centering
\begin{tabular}{|c|c|c|c|c|c|c|c|}
\hline
$p$   & $\rho_1$ & $\rho_2$ & $\tau$ & $\delta$ & $\mu_0$ & $n$ & $T$ \\ \hline
0.75 & 0.6      & 0.9      & 10     & 0.5      & 0.5   & 500 & 700    \\ \hline
\end{tabular}
\caption{This table reports the values assigned to the parameters in the numerical simulations.}
\label{tab_bench}
\end{table}

We vary $q$ from $0.5$ to $0.9$ in increments of $0.1$ (the case $q<0.5$ is
symmetric and thus omitted).  For each value of $q$, we repeat the simulation 100 times. At each run, we randomly determine the
initial partition of the population into types.\footnote{We use the randomization algorithm applied by \href{https://ccl.northwestern.edu/netlogo/bind/primitive/random.html}{NetLogo}. At the beginning of each simulation, each agent is independently assigned
one of the five types with equal probability.} 
The randomization of the initial
shares, combined with the number of repetitions, explores a broad set of starting
configurations, and the standard deviation of the long-run outcomes is small, so our results
do not depend on initial conditions. We conducted all simulations using NetLogo; for
technical explanations of the code, see the
\href{https://github.com/rrozzi-econ/An-evolutionary-analysis-on-narrative-selection/tree/main}{replication package}.

\subsection{Two Types and Uncertainty}\label{subsec_2t_unc}
Table~\ref{tab_twot_q<1_indip} shows the results for the LoM \say{Independent State} when the population consists only of standard and conformist agents, confirming P1: conformist agents have an evolutionary advantage over standard agents when $q<1$.

\subsection{Five Types and Uncertainty}
We now let all five types coexist. P6 predicts that conformist agents remain the dominant type. Tables \ref{tab_fivet_q<1_indip}-\ref{tab_fivet_q<1_self} report the results, one for each LoM, confirming P6.

\paragraph{Independent State.} Table \ref{tab_fivet_q<1_indip} reports the results. Conformist agents are the best-performing type for any value of $q$, with a share ranging from 38\% to 43\%, and every other type survives, with a share ranging from 8\% to 25\%: the population is steadily heterogeneous. The relative performance of the minority types follows the pattern of Proposition~\ref{prop_heuristics}. At the parameters of Table~\ref{tab_bench}, $D<0$ at each $q$, so over-confident agents polarize, and under-confident agents remain moderate. Accordingly, over-confident and standard agents perform well when $q$ is close to $1$, where extreme beliefs pay, whereas under-confident agents perform well when $q$ is close to $\frac{1}{2}$; the crossing occurs near $q=\frac{3}{4}$, as anticipated in Section~\ref{subsec_heuristics}.

Conformist agents outperform both because they switch between the two behaviors. By
Lemma~\ref{lem_constrained}, they dampen signals when more extreme than the crowd and amplify
them when more moderate, so they behave like under-confident agents when their belief runs
ahead of the anchor and like over-confident agents when it lags behind. This pays only if the
anchor is well calibrated, which is what P2 predicts and Table~\ref{tab:mean_variance_independent}
confirms: at the benchmark precision $p=0.75$, the mean absolute distance between the anchor and
$q$ is small at any $q$. Remark~\ref{rem_calibrationnew} gives the reason: selection recurrently concentrates beliefs around $q$.

\begin{table}[h!]
\centering
\begin{tabular}{llllll}
\hline
Type/$q$                & 0.5        & 0.6        & 0.7        & 0.8        & 0.9        \\ \hline
Under-confident      & 0.248    & 0.234    & 0.217    & 0.178     & 0.105     \\
Anti-conformist     & 0.101     & 0.111    & 0.112    & 0.094     & 0.077    \\
Over-confident       & 0.097     & 0.112    & 0.116     & 0.134      & 0.174     \\
Standard            & 0.153    & 0.159    & 0.16    & 0.172    & 0.215     \\
Conformist         & 0.4    & 0.384    & 0.395    & 0.421    & 0.429    \\ \hline
SD under-confident  & 0.022 & 0.021 & 0.026 & 0.024 & 0.016 \\
SD anti-conformist & 0.013 & 0.012 & 0.013 & 0.013 & 0.011 \\
SD over-confident   & 0.012 & 0.013 & 0.016 & 0.015 & 0.014 \\
SD standard        & 0.011  & 0.011 & 0.012  & 0.013  & 0.013      \\
SD conformist     & 0.038  & 0.037 & 0.034 & 0.033 & 0.022
\end{tabular}
\caption{This table reports the average and standard deviation of the population shares for the five types recorded in 100 simulations under the LoM ``Independent State'' using the benchmark values in Table \ref{tab_bench} for the parameters.}
\label{tab_fivet_q<1_indip}
\end{table}

\paragraph{Auto-correlated State.} Table~\ref{tab_fivet_q<1_auto} reports the results. Introducing path dependence leaves the performance of each type close to that under the LoM \say{Independent State}---see Table~\ref{tab_fivet_q<1_indip}---even though $q$ is now the long-run rather than the instantaneous probability that $\omega=A$. Thus, evolutionary performance depends on the overall uncertainty about the state rather than on the exact state dynamics.

\begin{table}[h!]
\centering
\begin{tabular}{llllll}
\hline
Type/$q$                & 0.5        & 0.6        & 0.7        & 0.8        & 0.9        \\ \hline
Under-confident      & 0.244     & 0.235    & 0.215    & 0.188    & 0.11    \\
Anti-conformist     & 0.1     & 0.114    & 0.121     & 0.104    & 0.085    \\
Over-confident       & 0.095    & 0.114    & 0.125    & 0.145     & 0.182    \\
Standard            & 0.151    & 0.159    & 0.164    & 0.181    & 0.216    \\
Conformist         & 0.41    & 0.377     & 0.375    & 0.382    & 0.407      \\ \hline
SD under-confident  & 0.022 & 0.025 & 0.032 & 0.027 & 0.016 \\
SD anti-conformist & 0.011 & 0.014 & 0.014 & 0.014 & 0.011 \\
SD over-confident  & 0.01 & 0.015 & 0.016  & 0.017 & 0.014 \\
SD standard        & 0.01 & 0.013 & 0.014 & 0.013 & 0.014 \\
SD conformist     & 0.037 & 0.043 & 0.041 & 0.032 & 0.024
\end{tabular}
\caption{This table reports the average and standard deviation of the population shares for the five types recorded in 100 simulations under the LoM ``Auto-correlated State'' using the benchmark values in Table \ref{tab_bench} for the parameters.}
\label{tab_fivet_q<1_auto}
\end{table}

\paragraph{Self-fulfilling State.} Table~\ref{tab_fivet_q<1_self} reports the results. Assuming that the probability of the state (but not evolutionary performance) depends on the average belief in the population does not alter our conclusions qualitatively. Comparing with the LoM \say{Independent State}---see Table~\ref{tab_fivet_q<1_indip}---conformist agents perform slightly worse, and under-confident agents better, at low $q$; raising $\delta$ to $1$, so that the probability of the state coincides with the average belief, strengthens both effects---see
Table~\ref{tab_fivet_q<1_self_d1}.
A higher $\delta$ makes the realization $\omega=A$ more likely, which in turn makes the average
belief more extreme. By Lemma~\ref{lem_constrained}, conformist agents then tend to find
themselves more moderate than the crowd and amplify their signals, behaving like over-confident
rather than under-confident agents. They therefore converge too quickly to an extreme belief, which raises their error when $q$ is low, leaving room for under-confident agents.

\begin{table}[h!]
\centering
\begin{tabular}{llllll}
\hline
Type/$q$                & 0.5        & 0.6        & 0.7        & 0.8        & 0.9        \\ \hline
Under-confident      & 0.262    & 0.244    & 0.225     & 0.189    & 0.108    \\
Anti-conformist     & 0.104     & 0.114    & 0.12     & 0.1    & 0.08    \\
Over-confident       & 0.1    & 0.114    & 0.125    & 0.14     & 0.172    \\
Standard            & 0.157    & 0.164    & 0.165    & 0.174    & 0.214    \\
Conformist         & 0.378    & 0.363    & 0.366     & 0.397     & 0.426    \\ \hline
SD under-confident  & 0.024 & 0.024 & 0.029 & 0.025 & 0.013  \\
SD anti-conformist & 0.011 & 0.013 & 0.015 & 0.015 & 0.01 \\
SD over-confident  & 0.011 & 0.014 & 0.013 & 0.014 & 0.013 \\
SD standard        & 0.013 & 0.011 & 0.012 & 0.012 & 0.012 \\
SD conformist     & 0.04 & 0.04 & 0.038 & 0.031 & 0.021
\end{tabular}
\caption{This table reports the average and standard deviation of the population shares for the five types recorded in 100 simulations under the LoM ``Self-fulfilling State'' using the benchmark values in Table \ref{tab_bench} for the parameters.}
\label{tab_fivet_q<1_self}
\end{table}

\subsection{Comparative Statics}\label{subsec_rob}

\paragraph{Higher Precision of Information}

P3 predicts that a higher precision reduces the advantage of conformist agents when $q$ is sufficiently high. We now assume $p=0.85$; Table~\ref{tab_ind_p085} reports the results for the LoM \say{Independent State}, confirming P3. Table~\ref{tab:mean_variance_independent} shows that raising $p$ from $0.75$ to $0.85$ increases the anchor's volatility for any $q$, since agents tend to receive the same information and their beliefs move together. A more volatile anchor is less reliable, which reduces the advantage of conformist agents. However, a higher $p$ also makes the other types' beliefs more extreme, worsening their evolutionary performance when uncertainty is high, thereby offsetting the conformist agents' loss from a more volatile anchor. Only anti-conformist agents gain for any $q$, as a more volatile anchor makes distance from the crowd easier to sustain.

\begin{table}[h!]
    \centering
    \caption{Mean absolute anchor distance and anchor volatility under the LoM ``Independent State''.}
    \label{tab:mean_variance_independent}
    \begin{tabular}{lccccc}
        \toprule
        & \multicolumn{5}{c}{$q$} \\
        \cmidrule(lr){2-6}
        & $0.5$ & $0.6$ & $0.7$ & $0.8$ & $0.9$ \\
        \midrule
        \multicolumn{6}{l}{$p=0.75$} \\
        $\frac{1}{T}\sum_{t=1}^{T}\left|\bar{\mu}_{t-1}(h_t)-q\right|$
            & 0.1091 & 0.0980 & 0.0786 & 0.0506 & 0.0262 \\
        $\mathrm{Var}\left(\bar{\mu}_{t-1}(h_t)\right)$
            & 0.0171 & 0.0131 & 0.0089 & 0.0041 & 0.0018 \\
        \addlinespace
        \multicolumn{6}{l}{$p=0.85$} \\
        $\frac{1}{T}\sum_{t=1}^{T}\left|\bar{\mu}_{t-1}(h_t)-q\right|$
            & 0.1558 & 0.1396 & 0.1101 & 0.0748 & 0.0396 \\
        $\mathrm{Var}\left(\bar{\mu}_{t-1}(h_t)\right)$
            & 0.0341 & 0.0252 & 0.0157 & 0.0073 & 0.0023 \\
        \bottomrule
    \end{tabular}
\end{table}

\paragraph{Less Heterogeneous Narratives}
We now reduce narrative heterogeneity, assuming $\rho_1=0.7$ and $\rho_2=0.8$, symmetric around
the true precision $p=0.75$. Table~\ref{tab_indep_drho01} reports the results for the LoM
\say{Independent State}. Consistent with P4, this reduction limits the extent to which conformist
agents can adjust their beliefs toward the anchor, worsening their performance; it also raises
$D$, which turns positive at high $q$, so under-confident beliefs are no longer held near
$\frac{1}{2}$ and their performance improves.

\paragraph{Alternative Initial Beliefs.} 
P5 predicts that conformist agents remain dominant when newborn agents adopt the average belief
instead of $\mu_0$. Table~\ref{tab_indep_mu0pop} confirms this, and also shows an improvement in the performance of anti-conformist agents: newborn agents no longer pull the average belief back toward $\mu_0$, so the anchor becomes more extreme, and holding a belief far from it pays more often.

\section{Conclusion}\label{sec_conc}
We study how agents employing different heuristics to select narratives (i.e., to
determine how to interpret information) perform from an evolutionary perspective. We take an
agent's accuracy in predicting the state probability as their evolutionary performance, assuming that more accurate types are more likely to survive.
Conformist agents, who anchor their interpretation to the population's average belief, are the dominant type in uncertain environments, but all other types hold a minority share: agents using different heuristics coexist in the long run.

In financial markets, the prevalence of conformism among investors need not imply that private information is discarded. A conformist investor still processes their information; what the anchor governs is only the precision they attribute to it. Since selection has already removed the investors whose beliefs were furthest from the truth, the anchor carries
information of its own, which conformism incorporates into the investor's reading of the news.
The remaining heuristics sort by predictability: investors who hold moderate beliefs
perform better when a firm's prospects are uncertain, and those who form extreme beliefs when they
are not.

Our results also have implications for information design. Imagine a politician wishing to
persuade voters. From a static perspective, the politician must account for the existence of
agents that process information differently. In the long run, information design could also shape
the evolutionary path of different types: the designer can shape information today to favor types
that are easier to persuade, creating a more favorable environment tomorrow. This interaction
between information design and population evolution is worth studying in future research.

\newpage

\normalsize
\begin{spacing}{1.0}

\end{spacing}

\appendix

\setcounter{table}{0}
\renewcommand{\thetable}{A\arabic{table}}

\setcounter{figure}{0}
\renewcommand{\thefigure}{A\arabic{figure}}

\clearpage

\section{Proofs}\label{app_proofs}

\subsection*{Proof of Lemma~\ref{lem_constrained}}
By equation~\eqref{Bayesian},
\begin{align*}
    \mu_{i,t}(a,h_{i,t},\rho)&=\frac{\rho\,\mu_{i,t-1}}{\rho\,\mu_{i,t-1}+(1-\rho)(1-\mu_{i,t-1})},\\[2pt]
    \mu_{i,t}(b,h_{i,t},\rho)&=\frac{(1-\rho)\,\mu_{i,t-1}}{(1-\rho)\,\mu_{i,t-1}+\rho\,(1-\mu_{i,t-1})},
\end{align*}
so that $\mu_{i,t}(b,h_{i,t},\rho)=\mu_{i,t}(a,h_{i,t},1-\rho)$. Differentiating,
\begin{equation*}
    \frac{\partial\mu_{i,t}(a,h_{i,t},\rho)}{\partial\rho}=\frac{\mu_{i,t-1}(1-\mu_{i,t-1})}{\big(\rho\,\mu_{i,t-1}+(1-\rho)(1-\mu_{i,t-1})\big)^2}>0,
\end{equation*}
so the posterior after the pro-attitudinal signal is strictly increasing in $\rho$ and, by the
identity above, the posterior after the counter-attitudinal signal is strictly decreasing in
$\rho$. Since $\rho_1<p<\rho_2$, it follows that $\mu_{i,t}(s_{i,t},h_{i,t},p)$ lies in the
interior of $\Xi_{i,t}(s_{i,t})$ after either signal. Finally, for any two attainable
posteriors $\mu'\neq\mu''$,
\begin{equation}\label{eq_twopoint}
    \big|\mu'-\bar{\mu}_{t-1}(h_t)\big|^2-\big|\mu''-\bar{\mu}_{t-1}(h_t)\big|^2=(\mu'-\mu'')\big(\mu'+\mu''-2\,\bar{\mu}_{t-1}(h_t)\big),
\end{equation}
so the closer of the two to $\bar{\mu}_{t-1}(h_t)$ is the one lying on the same side of their
midpoint $\frac{1}{2}(\mu'+\mu'')$, with indifference when $\bar{\mu}_{t-1}(h_t)$ equals the
midpoint.

\emph{Parts (i) and (ii).} The agent chooses between the two endpoints of
$\Xi_{i,t}(s_{i,t})$, whose midpoint is $\overline{M}(\mu_{i,t-1})$ if $s_{i,t}=a$ and
$\underline{M}(\mu_{i,t-1})$ if $s_{i,t}=b$. By equation \eqref{eq_twopoint}, they select the endpoint lying on the same side of that midpoint as $\bar{\mu}_{t-1}(h_t)$. By the monotonicity
established above, $\rho_2$ yields the higher endpoint after $a$ and the lower one after $b$,
which gives (i) and (ii).

\emph{Part (iii).} Suppose $\bar{\mu}_{t-1}(h_t)\notin\Xi_{i,t}(s_{i,t})$. The distance
$\big|\mu_{i,t}(s_{i,t},h_{i,t},\rho)-\bar{\mu}_{t-1}(h_t)\big|$ is then strictly monotone in
$\rho$, so it is strictly smaller at the $\rho\in\{\rho_1,\rho_2\}$ selected in (i) and (ii)
than at any $\rho\in(\rho_1,\rho_2)$, and $\rho_1<p<\rho_2$. \hfill$\blacksquare$

\subsection*{Proof of Proposition~\ref{prop_shrinkage}}

By equation~\eqref{eq_shrinkage} and the linearity of the expectation, recalling that $m_t$ is determined by $h_t$ and is therefore a constant under the conditional expectation,
\begin{equation*}
    E\big[\mu^C_{i,t}\mid h_t\big]=\lambda\,E\big[\tilde{\mu}_{i,t}\mid h_t\big]+(1-\lambda)\,m_t
    =\lambda\big(q+b_{i,t}\big)+(1-\lambda)\big(q+d_t\big)
    =q+\lambda b_{i,t}+(1-\lambda)d_t,
\end{equation*}
and, for the same reason,
\begin{equation*}
    \mathrm{Var}\big(\mu^C_{i,t}\mid h_t\big)=\mathrm{Var}\big(\lambda\,\tilde{\mu}_{i,t}\mid h_t\big)=\lambda^2\sigma^2_{i,t}.
\end{equation*}
Substituting into equation~\eqref{eq_biasvariance},
\begin{equation*}
    E\big[\big(q-\mu^C_{i,t}\big)^2\mid h_t\big]=\big(\lambda b_{i,t}+(1-\lambda)d_t\big)^2+\lambda^2\sigma^2_{i,t},
    \qquad
    E\big[\big(q-\tilde{\mu}_{i,t}\big)^2\mid h_t\big]=b_{i,t}^2+\sigma^2_{i,t},
\end{equation*}
and taking the difference yields equation~\eqref{eq_delta}.

Consider part (i). If $|d_t|\leq|b_{i,t}|$, the triangle inequality gives
\begin{equation*}
    \big|\lambda b_{i,t}+(1-\lambda)d_t\big|\leq\lambda|b_{i,t}|+(1-\lambda)|d_t|\leq\lambda|b_{i,t}|+(1-\lambda)|b_{i,t}|=|b_{i,t}|,
\end{equation*}
so that $\big(\lambda b_{i,t}+(1-\lambda)d_t\big)^2-b_{i,t}^2\leq0$. Since $\sigma^2_{i,t}>0$ and $\lambda\in[0,1)$, the remaining term $-(1-\lambda^2)\sigma^2_{i,t}$ is strictly negative, thus $\Delta_{i,t}<0$.

Consider part (ii). Since $\lambda\in[0,1)$, we have $1-\lambda^2>0$, so equation~\eqref{eq_delta} gives
\begin{equation*}
    \Delta_{i,t}<0
    \iff \big(1-\lambda^2\big)\sigma^2_{i,t}>\big(\lambda b_{i,t}+(1-\lambda)d_t\big)^2-b_{i,t}^2
    \iff \sigma^2_{i,t}>\bar{\sigma}^2_{i,t}. 
\end{equation*}
\hfill$\blacksquare$

\subsection*{Proof of Proposition~\ref{prop_configA}}
\emph{Part (i).} By equation~\eqref{eq_anchorbound}, $\bar{\mu}_{t}(h_{t+1})\leq\kappa$ at each turnover time. In configuration A, taking conditional expectations yields $m_t\leq\kappa$ at each turnover time; in configuration B, since the anchor is lagged, the
same holds at each time following a turnover time. At any such time, bounding $\tilde{\mu}_{i,t}$ by $1$ in equation~\eqref{eq_shrinkage},
\begin{equation*}
    \mu^C_{i,t}\;\leq\;\lambda+(1-\lambda)\,\kappa,\qquad\text{hence}\qquad 1-\mu^C_{i,t}\;\geq\;(1-\lambda)(1-\kappa),
\end{equation*}
and squaring gives $\epsilon_{i,t}\geq(1-\lambda)^2(1-\kappa)^2>0$. Since turnover times recur every $\tau$ periods, the inequality holds infinitely often and $\epsilon_{i,t}\not\to0$.
For the standard agent, at $q=1$ signals are i.i.d.\ with $P(s_{j,t}=a)=p>\frac{1}{2}$,
so the log-odds $\ln\frac{\mu^S_{j,t}}{1-\mu^S_{j,t}}$ are a random walk with increments $\pm\ln\frac{p}{1-p}$ and drift $(2p-1)\ln\frac{p}{1-p}>0$. By the strong law of large numbers, the log-odds diverge to $+\infty$ almost surely, hence $\mu^S_{j,t}\to1$ and
$\epsilon_{j,t}\to0$ almost surely.

\emph{Part (ii).} We prove $\mu^C_{i,t}\leq\mu^S_{j,t}$ by induction. At the common birth time, the two agents share the same prior, so the inequality reduces to equality. Suppose it holds at $t-1$. By equation~\eqref{Bayesian}, the posterior is strictly increasing in the prior, and the two agents receive the same signal; applying it to the weakly smaller prior therefore gives
\begin{equation*}
    \tilde{\mu}_{i,t}\;\leq\;\mu^S_{j,t}.
\end{equation*}
Together with the hypothesis $m_t\leq\mu^S_{j,t}$, equation~\eqref{eq_shrinkage} gives
\begin{equation*}
    \mu^C_{i,t}=\lambda\,\tilde{\mu}_{i,t}+(1-\lambda)\,m_t\;\leq\;\lambda\,\mu^S_{j,t}+(1-\lambda)\,\mu^S_{j,t}=\mu^S_{j,t},
\end{equation*}
which completes the induction. Since $q=1$, the error $\epsilon_{i,t}=(1-\mu_{i,t})^2$ is decreasing in the belief on $[0,1]$, so $\epsilon_{i,t}\geq\epsilon_{j,t}$ at each time and along every path. \hfill$\blacksquare$

\subsection*{Proof of Proposition~\ref{prop_interior}}
Write $\ell_{j,t}\coloneqq\ln\frac{\mu^S_{j,t}}{1-\mu^S_{j,t}}$ for the log-odds of agent $j$'s
belief. Since a standard agent updates with $\rho_{j,t}=p$, $\ell_{j,t}=\ell_{j,0}+\sum_{u=1}^{t}z_{j,u}$, where $z_{j,u}=\ln\frac{p}{1-p}$ if $s_{j,u}=a$
and $z_{j,u}=-\ln\frac{p}{1-p}$ otherwise. Under the LoM \say{Independent State}, the state is drawn independently in each period, so the
$z_{j,u}$ are i.i.d.\ with
\begin{equation*}
    P_a\coloneqq P(s_{j,u}=a)=qp+(1-q)(1-p)=(1-p)+q(2p-1)>\tfrac{1}{2},
\end{equation*}
since $q>\frac{1}{2}$ and $p>\frac{1}{2}$. Each $z_{j,u}$ therefore has expectation $(2P_a-1)\ln\frac{p}{1-p}>0$. Since $\ell_{j,t}/t=\ell_{j,0}/t+\frac{1}{t}\sum_{u=1}^{t}z_{j,u}$ and $\ell_{j,0}/t\to0$, the strong law of large numbers gives $\ell_{j,t}/t\to(2P_a-1)\ln\frac{p}{1-p}\eqqcolon c>0$ almost surely. Hence $\ell_{j,t}\to+\infty$ and $\mu^S_{j,t}\to1$, almost surely.

The bias and variance in equation~\eqref{eq_biasvariance} are conditional on $h_t$, which
determines $\mu^S_{j,t-1}$ and leaves only the period-$t$ signal random: $\mu^S_{j,t}$ takes two
values, both continuous in $\mu^S_{j,t-1}$ and equal to $1$ when $\mu^S_{j,t-1}=1$. Since
$\mu^S_{j,t-1}\to1$ almost surely, both converge to $1$, so $b_{j,t}\to1-q>0$ and
$\sigma^2_{j,t}\to0$, and equation~\eqref{eq_biasvariance} gives $E[\epsilon_{j,t}\mid
h_t]\to(1-q)^2$ almost surely.
\hfill$\blacksquare$

\subsection*{Proof of Remark~\ref{rem_calibrationnew}}
Let $\phi$ denote the density of the error $\epsilon$ on $[\underline{\epsilon},\bar{\epsilon}]$
at any turnover time, before selection, and let $w(\epsilon)\coloneqq1-g(\epsilon;\psi_t)$
denote the survival probability, which by equation~\eqref{survival} is nonnegative,
decreasing, and strictly positive for $\epsilon<\bar{\epsilon}$. Fix
$c\in[\underline{\epsilon},\bar{\epsilon}]$ and define
\begin{equation*}
    P\coloneqq\int_{c}^{\bar{\epsilon}}\phi(\epsilon)\,d\epsilon,\qquad
    A\coloneqq\int_{c}^{\bar{\epsilon}}w(\epsilon)\phi(\epsilon)\,d\epsilon,\qquad
    B\coloneqq\int_{\underline{\epsilon}}^{c}w(\epsilon)\phi(\epsilon)\,d\epsilon,
\end{equation*}
so that $P$ is the share of agents with error above $c$ before selection, while $A$ and $B$ are the masses of agents surviving with error above and below $c$. Their sum $A+B$ is the
surviving mass, and $\frac{A}{A+B}$ is the share of survivors with error above $c$. The claim is that $\frac{A}{A+B}\leq P$, or equivalently,
after cross-multiplication, that $A(1-P)\leq PB$.

Since $w$ is decreasing, $w(\epsilon)\leq w(c)$ for $\epsilon\geq c$ and
$w(\epsilon)\geq w(c)$ for $\epsilon\leq c$, so
\begin{equation*}
    A\;\leq\;w(c)\int_{c}^{\bar{\epsilon}}\phi(\epsilon)\,d\epsilon\;=\;w(c)\,P,
    \qquad
    B\;\geq\;w(c)\int_{\underline{\epsilon}}^{c}\phi(\epsilon)\,d\epsilon\;=\;w(c)\,(1-P).
\end{equation*}
Multiplying the first inequality by $1-P\geq0$ and the second by $P\geq0$,
\begin{equation*}
    A(1-P)\;\leq\;w(c)\,P(1-P)\;\leq\;P\,B,
\end{equation*}
which is the claim. As $c$ is arbitrary, the error among survivors is first-order
stochastically dominated by its pre-selection counterpart. \hfill$\blacksquare$

\subsection*{Proof of Proposition~\ref{prop_heuristics}}
Let $\mu_{i,t-1}\geq\frac{1}{2}$, so that $a$ is pro-attitudinal and $b$ counter-attitudinal.
An over-confident agent amplifies $a$ and dampens $b$, so their log-odds change by $+L_2$
with probability $P_a$ and by $-L_1$ with probability $1-P_a$, in expectation $\bar{D}$; an
under-confident agent does the opposite, obtaining $+L_1$ and $-L_2$, in expectation $D$. If
instead $\mu_{i,t-1}<\frac{1}{2}$, then $b$ is pro-attitudinal, and the same computation
gives $D$ for an over-confident agent and $\bar{D}$ for an under-confident one.

The log-odds $\ell_{i,t}\coloneqq\ln\frac{\mu_{i,t}}{1-\mu_{i,t}}$ are increasing in
$\mu_{i,t}$ and equal to zero at $\mu_{i,t}=\frac{1}{2}$. Hence, the belief moves toward $1$
when $\ell_{i,t}$ is expected to rise. It becomes more extreme when $\ell_{i,t}$ is expected
to rise and $\mu_{i,t-1}\geq\frac{1}{2}$, or to fall and $\mu_{i,t-1}<\frac{1}{2}$, and less
extreme in the two opposite cases. Since $q>\frac{1}{2}$ and $p>\frac{1}{2}$, we have $P_a>\frac{1}{2}$ and hence $\bar{D}>0$. If $D>0$, all four expected changes are positive, which proves (i). If $D<0$, an over-confident agent's expected change is positive when $\mu_{i,t-1}\geq\frac{1}{2}$ and negative when $\mu_{i,t-1}<\frac{1}{2}$, and an under-confident agent's the reverse, which proves (ii).

Finally, $D=P_a(L_1+L_2)-L_2$ is increasing in $P_a$, and $P_a$ is increasing in $q$ because
$p>\frac{1}{2}$ and in $p$ because $q>\frac{1}{2}$; and $D$ is increasing in $L_1$ and decreasing in $L_2$, both of which move $D$ down as $\rho_2-\rho_1$ widens.
\hfill$\blacksquare$

\section{Additional Tables}

\begin{table}[h!]
\centering
\begin{tabular}{lll|l}
\hline
$q$   & Standard & Conformist & SD        \\ \hline
0.5 & 0.325  & 0.675     & 0.033 \\
0.6 & 0.35  & 0.65     & 0.038 \\
0.7 & 0.373  & 0.627     & 0.045 \\
0.8 & 0.393  & 0.607     & 0.034 \\
0.9 & 0.415  & 0.585     & 0.022
\end{tabular}
\caption{This table reports the average and standard deviation of the population shares of standard and conformist agents recorded in 100 simulations under the LoM ``Independent State'', using the benchmark values in Table \ref{tab_bench} for the parameters. Note that as the average shares of Standard and Conformist agents sum to $1$, their standard deviation is the same.}
\label{tab_twot_q<1_indip}
\end{table}

\begin{table}[h!]
\centering
\begin{tabular}{llllll}
\hline
Type/$q$                &  0.5        &  0.6        &  0.7        &  0.8        &  0.9        \\ \hline
Under-confident      & 0.282    & 0.262    & 0.238     & 0.192    & 0.107    \\
Anti-conformist     & 0.115    & 0.128    & 0.123    & 0.104     & 0.08    \\
Over-confident       & 0.11     & 0.125    & 0.13    & 0.144    & 0.174     \\
Standard            & 0.158    & 0.165    & 0.168     & 0.178     & 0.217    \\
Conformist         & 0.335    & 0.321    & 0.342     & 0.382    & 0.422    \\ \hline
SD under-confident  & 0.024 & 0.025 & 0.024 & 0.026 & 0.015 \\
SD anti-conformist & 0.016 & 0.016  & 0.014 & 0.014 & 0.011 \\
SD over-confident  & 0.014 & 0.015 & 0.015 & 0.017  & 0.013  \\
SD standard        & 0.011 & 0.012 & 0.011 & 0.012 & 0.014 \\
SD conformist     & 0.042 & 0.039 & 0.036 & 0.032 & 0.025
\end{tabular}
\caption{This table reports the average and standard deviation of the population shares for the five types recorded in 100 simulations under the LoM ``Self-fulfilling State'' with $\delta=1$, using the benchmark values in Table \ref{tab_bench} for the parameters.}
\label{tab_fivet_q<1_self_d1}
\end{table}

\begin{table}[h!]
\centering
\begin{tabular}{llllll}
\hline
Type/$q$                & 0.5        & 0.6 & 0.7        & 0.8        & 0.9        \\ \hline
Under-confident      & 0.242    & 0.213                  & 0.205    & 0.204    & 0.155    \\
Anti-conformist     & 0.109    & 0.148                  & 0.131    & 0.126    & 0.11    \\
Over-confident       & 0.086     & 0.098                    & 0.106    & 0.132    & 0.172    \\
Standard            & 0.133 & 0.138               & 0.155 & 0.162 & 0.201 \\
Conformist          & 0.433    & 0.404                   & 0.405    & 0.378    & 0.364    \\ \hline
SD under-confident  & 0.1 & 0.103               & 0.1 & 0.085 & 0.059 \\
SD anti-conformist & 0.074 & 0.132               & 0.079 & 0.102  & 0.048 \\
SD over-confident   & 0.036 & 0.048               & 0.055 & 0.058 & 0.066  \\
SD standard        & 0.066 & 0.064               & 0.1 & 0.065  & 0.063 \\
SD conformist      & 0.174 & 0.186               & 0.168 & 0.168 & 0.171
\end{tabular}
\caption{This table reports the average and standard deviation of the population shares for the five types recorded in 100 simulations under the LoM ``Independent State'' with $p=0.85$, using the benchmark values in Table \ref{tab_bench} for the other parameters.}
\label{tab_ind_p085}
\end{table}

\begin{table}[h!]
\centering
\begin{tabular}{llllll}
\hline
Type/$q$                & 0.5        & 0.6 & 0.7        & 0.8        & 0.9        \\ \hline
Under-confident      & 0.220 & 0.207                    & 0.200 & 0.202 & 0.172 \\
Anti-conformist     & 0.150 & 0.171                    & 0.169 & 0.185 & 0.186 \\
Over-confident       & 0.125 & 0.137                    & 0.148 & 0.165 & 0.161 \\
Standard            & 0.171 & 0.175                    & 0.181 & 0.191 & 0.184 \\
Conformist          & 0.336 & 0.311                    & 0.303 & 0.258 & 0.297 \\ \hline
SD under-confident  & 0.078 & 0.072                    & 0.071 & 0.056 & 0.074 \\
SD anti-conformist & 0.082 & 0.106                    & 0.113 & 0.097 & 0.084 \\
SD over-confident   & 0.045 & 0.053                    & 0.055 & 0.048 & 0.071 \\
SD standard        & 0.053 & 0.062                    & 0.068 & 0.052 & 0.057 \\
SD conformist      & 0.178 & 0.151                    & 0.185 & 0.134 & 0.198
\end{tabular}
\caption{This table reports the average and standard deviation of the population shares for the five types recorded in 100 simulations under the LoM ``Independent State'' with $\rho_1=0.7$ and $\rho_2=0.8$, using the benchmark values in Table \ref{tab_bench} for the other parameters.}
\label{tab_indep_drho01}
\end{table}

\begin{table}[h!]
\centering
\begin{tabular}{llllll}
\hline
Type/$q$                & 0.5        & 0.6 & 0.7        & 0.8        & 0.9        \\ \hline
Under-confident      & 0.207 & 0.216 & 0.214 & 0.188 & 0.104 \\
Anti-conformist     & 0.124 & 0.134 & 0.131 & 0.150 & 0.177 \\
Over-confident        & 0.084 & 0.120 & 0.139 & 0.151 & 0.163 \\
Standard            & 0.128 & 0.155 & 0.160 & 0.158 & 0.172 \\
Conformist         & 0.457 & 0.375 & 0.356 & 0.353 & 0.384 \\ \hline
SD under-confident  & 0.097 & 0.070 & 0.063 & 0.074 & 0.046 \\
SD anti-conformist & 0.131 & 0.108 & 0.065 & 0.102 & 0.105 \\
SD over-confident   & 0.044 & 0.042 & 0.042 & 0.052 & 0.074 \\
SD standard        & 0.053 & 0.067 & 0.035 & 0.052 & 0.064 \\
SD conformist     & 0.180 & 0.156 & 0.141 & 0.155 & 0.196
\end{tabular}
\caption{This table reports the average and standard deviation of the population shares for the five types recorded in 100 simulations under the LoM ``Independent State'', using the benchmark values in Table \ref{tab_bench} for the parameters, and assuming that each agent is born with the average belief in the population.}
\label{tab_indep_mu0pop}
\end{table}

\clearpage

\end{document}